\documentclass{llncs}
\usepackage{amsmath,amssymb}
\usepackage{graphicx}
\usepackage{xcolor}
\usepackage{cite}
\usepackage{booktabs}
\usepackage{tikz}
\usetikzlibrary{arrows.meta, positioning, shapes.multipart}
\usepackage{siunitx}
\DeclareSIUnit{\gas}{gas}
\newcommand{\mgs}{\si{\mega\gas\per\second}}
\usepackage[hidelinks]{hyperref}
\newcommand{\Description}[1]{}

\begin{document}

\title{Fuji: The Shape, Composition, and Spread of Demand Bursts on Robinhood Chain}
\author{Pranay Anchuri}
\institute{Offchain Labs\\\email{panchuri@offchainlabs.com}}
\maketitle

\begin{abstract}
Demand on blockchain-based marketplaces arrives in bursts and at a high rate. On these
platforms the data are completely visible, which makes it feasible to compare the bursty
demand against the usual load. The congestion manifests as an increased demand for gas, the
primary resource a blockchain sells. Existing studies consider how to price the gas. The demand
itself, and what it is made of during bursts, has not been measured. We measure it on Robinhood Chain, a
layer-2 blockchain that quickly became a venue for trading tokenized equities and memecoins.
We collected the data from every block of its first three active months and more than 50
million Uniswap swaps that occurred during bursty periods. We find that a burst trades about as many pools as a quiet period, mostly different
ones, and trades them harder, on top of elevated demand that accumulated for hours. With the bursts aligned on their peaks, the median peak is 2.5 times the usual load. Both the rise and the fall
of the demand are described by the sum of two exponential functions with different timescales,
a slow one of hours and a fast one of minutes. Against a matched quiet
period, a burst has 1.05 times the pools, 1.40 times the trades, and 1.21 times the senders. We
use two models to explain the arrival of bursts: externally triggered cascades and
self-excitation. The former fits the data relatively better. Inside a burst, the entry of each
trader into a pool is followed for about 24 seconds by more entries than the existing traders
predict. Whether this is because of imitation or because of a shared signal is left
open.
\end{abstract}

\section{Introduction}

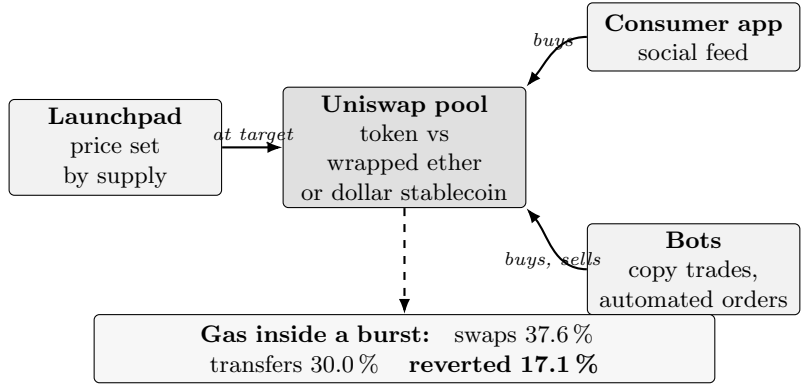
\begin{figure}[t]
\centering
\begin{tikzpicture}[
  box/.style={draw, rounded corners=2pt, align=center, font=\footnotesize,
              minimum height=9mm, inner sep=3pt},
  app/.style={box, fill=black!5, text width=26mm},
  pool/.style={box, fill=black!12, text width=30mm},
  ar/.style={-{Latex[length=2mm]}, thick},
  lbl/.style={font=\scriptsize\itshape, inner sep=1pt}
]
\node[app] (lp) {\textbf{Launchpad}\\price set by supply};
\node[pool, right=8mm of lp] (pool) {\textbf{Uniswap pool}\\token vs wrapped ether\\or dollar stablecoin};
\node[app, above right=2mm and 8mm of pool] (feed) {\textbf{Consumer app}\\social feed};
\node[app, below right=2mm and 8mm of pool] (bots) {\textbf{Bots}\\copy trades,\\automated orders};
\draw[ar] (lp) -- node[lbl, above] {at target} (pool);
\draw[ar] (feed.west) to[out=180, in=50] node[lbl, above] {buys} (pool.north east);
\draw[ar] (bots.west) to[out=180, in=-50] node[lbl, below] {buys, sells} (pool.south east);
\node[box, below=14mm of pool, text width=80mm, fill=black!3] (gas)
 {\textbf{Gas inside a burst:}\quad swaps 37.6\,\% \;\; transfers 30.0\,\% \;\;
  \textbf{reverted 17.1\,\%}};
\draw[ar, dashed] (pool) -- (gas);
\end{tikzpicture}
\caption{How a trade reaches the chain. On the launchpad, new tokens are sold at a price that
depends on their supply. They move to Uniswap pools once purchases reach a fixed target.
Traders then trade them through a social-feed app or through bots. The bottom bar shows the
share of gas by transaction type inside burst intervals (Section~\ref{sec:comp}).}
\label{fig:flow}
\Description{Block diagram: launchpad feeds a Uniswap pool; a consumer app and bots trade against the pool; a bar below lists gas shares inside a burst.}
\end{figure}

On one afternoon, the gas usage on Robinhood Chain was several times above the usual for one minute and then it stayed at an elevated level for more than one hour afterward. Figure~\ref{fig:windows} shows the gas usage during 82 minutes of this burst activity. This period contained \num{138609} swaps on Uniswap, touching \num{1411} different pools. The users on this chain include both retail users and automated trading bots. They primarily trade memecoins, which are initially sold at a price that depends on their supply. These tokens are traded along with equities and funds such as NVDA,
GME, and SPY. The users buy these tokens both through a social feed that shows what others are buying and through bots that copy others’ trades (Figure~\ref{fig:flow}). The periods of elevated demand on Robinhood Chain can be described by a slow and gradual rise and fall around an intense peak.

The research question we study in this work is how the resource demand concentrates both in time and across assets in an online marketplace, Robinhood Chain. The resource we study here is the gas, the unit of blockchain computation.
A public blockchain makes the link between congestion, participation, and trading value observable. The transactions happening on the blockchain are organized into blocks. The block records the load and the transaction records the trades inside it in an immutable public database. Records like these have been used to study user preferences in a range of fields. Wu and Huberman~\cite{wu07} record
votes on news stories, and Salganik et al.~\cite{salganik06} record song choices in an
experimental market. Different from our case, these platforms record cases without the platform’s load. On the blockchain, the platform load can be connected to the count and value of trades. Moreover, the gas price adjusts according to the demand. Existing studies look at the fee mechanisms ~\cite{roughgarden21,chungshi23,reijsbergen21,liu22} in great detail, but, critically, they do not model the demand and the activity during periods of elevated demand.

Measuring the demand for the gas is different from measuring its price. The same fee level can come from user activity spread throughout the day or concentrated into bursts. Moreover, a burst that is caused by many trading pools has different implications from a burst concentrated in a few. The spacing between the bursts determines what the current burst says about the next one. As demand bursts can overlap, a burst cannot be easily identified by comparing it to nearby (in time) demand.

We collected the complete block record of Robinhood Chain and every Uniswap swap that happened during periods of demand bursts (Section~\ref{sec:data}). We identify these bursts using a moving average version of the gas usage. But the bursts are characterized by drawing a fixed window around the peak minute of the burst. In addition to identifying the bursty periods, we measure the trades and the value during them. We also follow the traders and the pools through a burst. These measurements of the Uniswap trades on Robinhood Chain support the following contributions.

\begin{enumerate}

\item \textbf{Shape of demand bursts.} We develop a two-exponential model to describe the profile of the median burst. When the model is fitted to the observed data, the slow timescale is about 60 times the fast one. It is about 40 times after controlling for the daily cycle (Section~\ref{sec:shape}). The control time period (24 hours prior) accounts for at most 38 percent of the slow rise before the peak. The peak of the demand is 2.5 times the usual and 1.6 times the demand 20 minutes away.

\item \textbf{Composition of bursts.} A burst trades about as many pools as a quiet period, but largely different ones, and trades them harder. There are more trades from more participants (Section~\ref{sec:comp}).
The trades account for the highest gas usage, and the gas spent on reverting transactions roughly doubles inside the demand bursts. The leading pool by value holds about a sixth of a burst's value, and the leading pool by trades about a sixth of its trades. Pools that are absent at the start attract more trades than value by the end of the burst.

\item \textbf{Clustering between and within bursts.} We fit the arrival of bursts using two different models: cascades and self-excitation (Section~\ref{sec:when}). The cascade fits relatively better. But the self-excitation model cannot be discarded. Inside the bursts, the entries of traders cluster. This effect decays over about 24 seconds and is stronger in pools with more swaps (Section~\ref{sec:spread}).

\end{enumerate}

\section{Data}\label{sec:data}

General-purpose blockchains such as Ethereum (and the rollup/layer-2 blockchains that are built on it) primarily sell gas. Most resources on the chain are denominated in gas units. Each transaction consumes some amount of gas, and each block at most a fixed amount of gas. The blockchain prices the gas based on the demand for it ~\cite{roughgarden21}. Robinhood Chain, which we study in this work, is a rollup that produces about ten blocks per second. The chain is public. So every block and the transactions in it are observable. We use the on-chain data from 27
June to 28 September 2026, the chain’s first three months of continuous
operation.

\subsection{Definitions}

Each block on the chain has a \texttt{gasUsed} field, which is the sum of the gas used by all transactions in the block (Figure~\ref{fig:records}). For each second, we sum the \texttt{gasUsed} of all blocks with that timestamp. Let $g(m)$ be the average of the gas used per second during minute $m$. The \emph{gas series} $G = (g(m))_m$ is the sequence of these averages over the entire dataset. The median of the gas series, $g_u$, is the chain’s \emph{usual level},
\begin{equation}
g_u \;=\; \operatorname{median}(G) \;=\; 16.1~\mgs ,
\label{eq:usual}
\end{equation}
and every multiple reported in this paper is a multiple of $g_u$.

For every minute $m$, we define the \emph{baseline} $b(m)$ as the median of $g$ over the 24 hours centered on $m$.
We also define $\bar g(m)$ as the 30-minute average of $g$ centered on $m$. A minute $m$ is considered \emph{busy} when
\begin{equation}
g(m) \;>\; 1.5\,b(m) \quad\text{and}\quad g(m) \;>\; b(m) + 5~\mgs .
\label{eq:def}
\end{equation}

The second condition is needed to account for the fact that $b(m)$ approaches zero during periods of low activity.
We derive the following subsequences from the gas series $G$ and the busy minutes.
\begin{align}
I &= \text{maximal run}\ge 5\ \text{busy minutes} , \label{eq:interval}\\
p(I) &= \arg\max\nolimits_{m\in I}\, g(m) , \label{eq:peak}\\
A(I) &= \{\, m : |m - p(I)| \le 360 \,\} , \label{eq:align}\\
W &= \text{maximal run}\ge 30\,\text{min},\ \bar g > 19\,\mgs , \label{eq:window}\\
C(W) &= W \cap \textstyle\bigcup_I I , \qquad M(W) = W \setminus C(W) . \label{eq:core}
\end{align}
\begin{itemize}

\item \textbf{Burst interval} $I$ (\ref{eq:interval}): a contiguous period of busy minutes. Its \emph{peak
minute} $p(I)$ (\ref{eq:peak}) is the minute of highest gas usage. As the baseline $b(m)$ is a rolling measure, the burst interval only locates a burst.

\item \textbf{Alignment window} $A(I)$ (\ref{eq:align}): a period of twelve hours centered on the peak
        minute. We compare the bursts by aligning them on their peak minutes. From the aligned bursts, we derive a profile of the burst by computing the median gas usage across all the bursts (Section~\ref{sec:shape}).

  \item \textbf{Burst window} $W$ (\ref{eq:window}): a contiguous period of minutes in which $\bar g$, a smoothed version of $G$, is above a fixed threshold. The period is further split into a \emph{core} and a \emph{margin}. The \emph{core} is the set of minutes that also lie in some burst interval. The remaining minutes are considered the \emph{margin}. We exclude windows that are longer than twelve hours.

\end{itemize}

\begin{figure}[t]
\centering
\resizebox{\textwidth}{!}{\begin{tikzpicture}[x=0.70mm, y=5.2mm, font=\scriptsize]
\draw[->, gray!60] (-2,0) -- (88,0) node[right, black, font=\tiny] {min from 13:17};
\draw[gray!60] (-2,0) -- (-2,3.6);
\foreach \y in {1,2,3} \draw[gray!60] (-4,\y) -- (-2,\y)
  node[left, black, xshift=-1.4mm, font=\tiny] {\y$\times$};
\foreach \a/\b in {13/20, 45/59, 64/70} \fill[black!13] (\a,0) rectangle (\b,3.6);
\draw[thick, black!75] plot coordinates {
(0,1.01) (1,1.03) (2,1.43) (3,1.11) (4,1.16) (5,1.04) (6,1.01) (7,1.11) (8,1.23) (9,1.24)
(10,1.45) (11,1.52) (12,1.20) (13,3.17) (14,3.10) (15,2.30) (16,1.84) (17,1.52) (18,2.03)
(19,2.18) (20,1.27) (21,1.17) (22,1.37) (23,1.17) (24,1.32) (25,1.27) (26,1.33) (27,1.59)
(28,1.35) (29,1.49) (30,1.74) (31,1.50) (32,1.28) (33,1.20) (34,1.21) (35,1.07) (36,1.01)
(37,1.36) (38,1.14) (39,1.11) (40,1.40) (41,1.05) (42,1.10) (43,1.44) (44,1.21) (45,1.85)
(46,1.89) (47,1.85) (48,2.27) (49,1.98) (50,1.48) (51,1.66) (52,1.76) (53,1.81) (54,1.79)
(55,1.50) (56,1.54) (57,1.45) (58,1.77) (59,1.35) (60,1.91) (61,1.43) (62,1.46) (63,1.23)
(64,1.71) (65,1.76) (66,1.60) (67,1.88) (68,1.73) (69,1.50) (70,1.31) (71,1.10) (72,1.07)
(73,1.28) (74,1.34) (75,1.19) (76,1.14) (77,1.31) (78,1.09) (79,0.91) (80,1.00) (81,1.03)};
\fill[black] (13,3.17) circle (0.5mm);
\node[font=\tiny, anchor=west] at (15,3.37) {peak minute};
\node[font=\tiny, anchor=south] at (16.5,2.37) {core};
\node[font=\tiny, anchor=south] at (52,2.34) {core};
\node[font=\tiny, anchor=south, black!60] at (32,0.59) {margin};
\draw[|-|, thick] (0,-1.0) -- (82,-1.0)
  node[midway, below, font=\tiny] {burst window, 82 min};
\draw[thick, black!55] (6,-2.3) -- (76,-2.3);
\node[font=\tiny, fill=white, inner sep=1pt] at (41,-2.3) {alignment window, 6 h either side};
\foreach \s in {-1,1} \foreach \d in {0,2}
  \draw[thick, black!55] ({41+\s*(35-\d)},-2.9) -- ({41+\s*(36.5-\d)},-1.7);
\draw[->, thick, black!55] (3,-2.3) -- (-2,-2.3);
\draw[->, thick, black!55] (79,-2.3) -- (86,-2.3);
\end{tikzpicture}}
\caption{A burst window of 82 minutes, with gas in multiples of the usual level, annotated
with the subsequences defined in Section~\ref{sec:data}. Shaded bands are burst intervals. The
alignment window extends beyond the plot.}
\label{fig:windows}
\Description{Minute-level gas trace with burst intervals shaded inside the burst window. The alignment window extends beyond the trace around a peak minute.}
\end{figure}
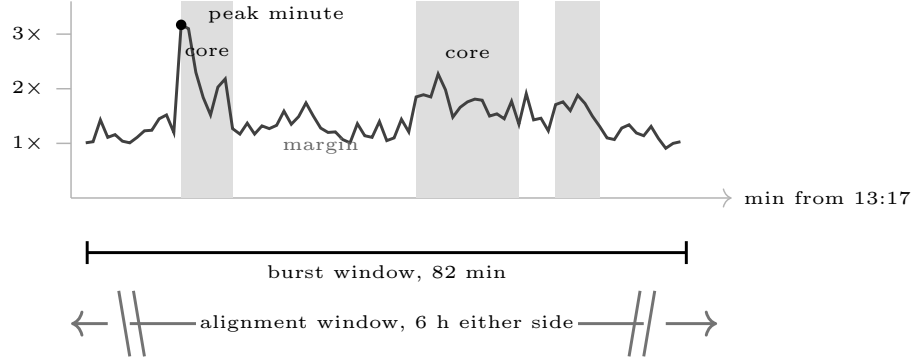

Table~\ref{tab:data} gives a summary of the above subsequences, and Figure~\ref{fig:windows} shows the three windows for one burst. The excluded windows (because of the twelve-hour threshold) account for 48\% of all minutes above the threshold in (\ref{eq:window}). Therefore, the window-level results in this paper only describe bursts.

\begin{table}[t]
\caption{Summary of the data and the subsequences derived from it (Section~\ref{sec:data}). Shares
are computed over all minutes and all gas usage in the three months.}
\label{tab:data}
\centering
\small
\begin{tabular}{lr}
\toprule
usual level $g_u$, median of $G$ & 16.1 \mgs \\
\midrule
burst intervals & 656 \\
\quad median length & 8 min \\
\quad share of minutes, share of gas & 6.1\%, 12.2\% \\
\quad with an alignment window inside the record & 654 \\
\quad sampled with a matched control & 150 \\
\midrule
burst windows detected & 214 \\
\quad median length & 77 min \\
\quad kept, at most twelve hours long & 200 \\
\quad with a core & 133 \\
\quad burst intervals lying inside a kept window & 437 \\
\midrule
Uniswap \texttt{Swap} events inside the 200 windows & \num{53618078} \\
\bottomrule
\end{tabular}
\end{table}

\subsection{Trading Data}\label{sec:trading}

The gas series describes the shape of the demand. But the composition of the demand requires inspecting the transactions within. On Robinhood Chain, the main application is Uniswap, a set of smart contracts (programs stored on the chain). In the burst intervals, the Uniswap trades account for 37.6\% of the gas (Section~\ref{sec:comp}). Each \emph{pool} on Uniswap holds two tokens and exchanges one for the other at an exchange price derived from the pool's state~\cite{angeris20,schar21}. At least one of these two tokens is a \emph{quote asset}, wrapped ether or a stablecoin. Some of the pools on Uniswap are \emph{routing pools}, where both tokens are quote assets. These pools route trades between other pools. We exclude them from the analysis. Each exchange on Uniswap emits a \texttt{Swap} log. A \emph{trader} is the address that signed and sent the transaction (its \texttt{from} field).

We collected swap events in two sets. First, inside the 200 burst windows we recorded the number of swaps per pool and per minute. For the twenty minutes around each window's peak, we further extracted the pool, block number, and trader of every swap. Section~\ref{sec:comp} uses the counts and Section~\ref{sec:spread} the swap-level records. Second, we recorded all the fields, including the amount field, from the 656 burst intervals. This set is used to price trades in Section~\ref{sec:comp}. Throughout the paper, the \emph{control} of a burst interval is the period of the same length but 24 hours earlier. We collected controls for a random sample of 150 intervals where the control does not overlap with a burst. The burst-to-control comparisons in Section~\ref{sec:comp} are based on this sample.

\begin{figure}[t]
\centering
\resizebox{0.7\textwidth}{!}{\begin{tikzpicture}[
  cls/.style={rectangle split, rectangle split parts=2, draw, rounded corners=1pt,
              rectangle split part fill={black!10, white}, font=\scriptsize, align=left,
              inner sep=3pt, text width=#1},
  node distance=2mm
]
\node[cls=26mm, anchor=north west] (b) at (0,0) {\textbf{Block header}
  \nodepart{two} block: 5225414\\ timestamp:\\ \quad 2026-07-09 12:42:00\\ \texttt{gasUsed}: \num{1405788}};
\node[cls=19mm, anchor=north west] (e) at ([xshift=2mm]b.north east) {\textbf{Swap event}
  \nodepart{two} block: 5225414\\ pool: 0x3eb8\ldots\\ tx from: 0x4433\ldots};
\node[cls=28mm, anchor=north west] (a) at ([xshift=2mm]e.north east) {\textbf{Swap with amounts}
  \nodepart{two} pool: 0xa660\ldots\\ minute: 2026-07-08 03:23\\ swaps: 3\\
  token0 in, out: 1.98e17, 1.31e17\\ token1 in, out: 3.64e25, 5.25e25};
\end{tikzpicture}}
\caption{Data fields and an example for each of the three record types used in the paper
(Section~\ref{sec:data}). Token amounts are in each token's smallest unit, before conversion to
dollars. Addresses are abbreviated.}
\label{fig:records}
\Description{Three class-diagram boxes: block header, swap event, and swap with amounts, each listing field names with example values.}
\end{figure}

\section{Shape of the Demand Burst}\label{sec:shape}

\begin{figure}[t]
\centering
\includegraphics[width=\columnwidth]{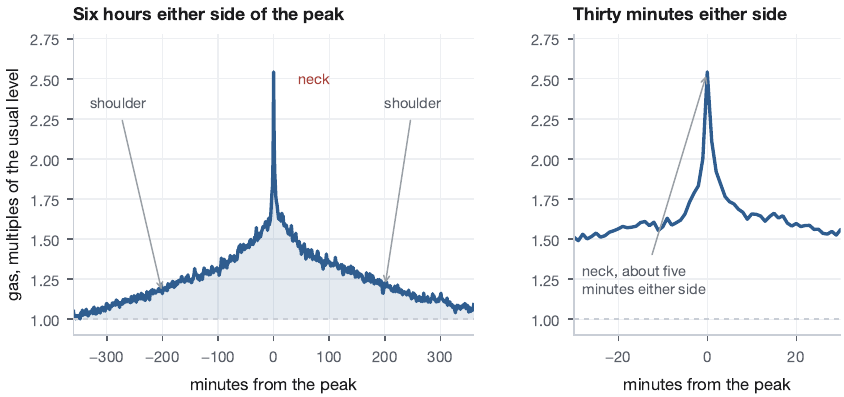}
\caption{The median alignment window $Z(t)$ derived from 654 bursts, in multiples of the usual
level. The left panel shows the full window and the right panel thirty minutes either side of
the peak. The shoulder and the neck are marked.}
\label{fig:fuji}
\Description{Median gas profile aligned on burst peaks. The broad view shows the shoulder, and the close view shows the sharp peak.}
\end{figure}

In this section, we model the shape of the demand bursts observed in the data. Section~\ref{sec:data} defines the alignment window of a burst interval as the twelve-hour time period around its peak minute (Equation~\ref{eq:align}).
From this, the \emph{median alignment window} is defined as
\begin{equation}
Z(t) \;=\; \operatorname{median}_{I}\, g\bigl(p(I) + t\bigr), \qquad |t| \le 360 ,
\label{eq:median}
\end{equation}
where the median is over the 654 bursts and $p(I)$ is the peak
minute of Equation~\ref{eq:peak}. Figure~\ref{fig:fuji} shows it, and Table~\ref{tab:profile}
gives the median gas usage at different offsets from the peak. At the peak, the interquartile range is 2.04 to 3.45
times the usual gas usage. Visual inspection of the median alignment window suggests two things. One, the shape is
nearly symmetric about the peak. Two, before the peak there is a slow rise followed by a rapid ascent confined to a few minutes. After the peak, there is a rapid descent confined to a few minutes followed by a slow fall. We call the slow part the \emph{shoulder} and the rapid part the \emph{neck}.\footnote{Mount Fuji has a similar profile, a long gentle slope and a steep summit. This similarity gives the paper its name.} The shape of the median is also observed at other percentiles (Figure~\ref{fig:aligned}, Appendix~\ref{app:det-shape}).

\begin{table}[t]
\caption{The median alignment window at selected offsets from the peak (multiples of the usual
level). The second row is the control profile, 24 hours earlier (Section~\ref{sec:cycle}).}
\label{tab:profile}
\centering
\small
\begin{tabular}{lrrrrrrr}
\toprule
& \multicolumn{3}{c}{before the peak} & & \multicolumn{3}{c}{after the peak} \\
\cmidrule(r){2-4}\cmidrule(l){6-8}
& 6 h & 2 h & 5 min & peak & 5 min & 2 h & 6 h \\
\midrule
burst & 1.05 & 1.30 & 1.66 & \textbf{2.54} & 1.73 & 1.32 & 1.09 \\
a day earlier & 1.07 & 1.28 & 1.30 & 1.37 & 1.36 & 1.27 & 1.08 \\
\bottomrule
\end{tabular}
\end{table}

\subsection{Burst Model}

The shoulder and neck of the demand burst suggest the use of a slow and a fast exponential function. Each flank of the median window is fitted separately to
\begin{equation}
\frac{Z(t)}{g_u} \;=\; c \;+\; A_{\mathrm{fast}}\,e^{-|t|/\tau_{\mathrm{fast}}}
\;+\; A_{\mathrm{slow}}\,e^{-|t|/\tau_{\mathrm{slow}}} ,
\label{eq:fuji}
\end{equation}
where $g_u$ is the usual level of Equation~\ref{eq:usual}, and $c$ is the constant to which the profile decays as $|t| \gg \tau_{\mathrm{slow}}$. Since $\tau_{\mathrm{slow}}$ is comparable to the fitted range of 360 minutes, $c$ is extrapolated and not observed in the data. Each $A, \tau$ pair is the amplitude and timescale of one exponential. The five parameters are fit using nonlinear least squares over offsets of 1 to 360 minutes. We also verify the need for two exponentials by fitting the flanks with a single exponential.
Table~\ref{tab:fit} gives both of these fits.

Three results follow from the fit. First, the two timescales differ by a factor of about 60 (80 on the rising flank and 47 on the falling flank). The timescales for the fast exponential are 2.7 minutes and 4.2 minutes for the rise and fall, respectively. Similarly, the slow timescales are 218.5 minutes and 199.2 minutes. Moreover, the slow term and the fast term are relatively less certain on the rising flank and the falling flank, respectively (Appendix~\ref{app:det-shape}). Second, a single exponential does not describe the profile. Its error is about 1.6 times that of the two-exponential fit on both flanks (Table~\ref{tab:fit}). Third, the peak is 2.54 times the chain's usual level and 1.85 times the level at the same time a day earlier (Table~\ref{tab:profile}). But it is only 1.6 times the level twenty minutes from the peak. This is because the slow term has already lifted the usage higher. The fit is weakest at the top, leaving around 40\% of the final rise unexplained. Therefore, the peak is sharper than suggested by the model.

\begin{table}[t]
\caption{Fit of Equation~\ref{eq:fuji} to each flank of the median alignment window. The
timescales are in minutes and the levels are multiples of the usual level. The bottom part shows
the slow timescale after controlling for the daily cycle (Section~\ref{sec:cycle}) and under a
6-hour detection baseline (Appendix~\ref{app:det-shape}).}
\label{tab:fit}
\centering
\small
\begin{tabular}{lrr}
\toprule
& rising flank & falling flank \\
\midrule
$\tau_{\mathrm{fast}}$ & 2.7 & 4.2 \\
$\tau_{\mathrm{slow}}$ & 218.5 & 199.2 \\
floor $c$ & 0.90 & 0.96 \\
$A_{\mathrm{fast}}$ & 0.57 & 0.54 \\
$A_{\mathrm{slow}}$ & 0.70 & 0.65 \\
RMSE, two exponentials & 0.020 & 0.023 \\
RMSE, one exponential & 0.033 & 0.038 \\
$\tau_{\mathrm{slow}}/\tau_{\mathrm{fast}}$ & 80 & 47 \\
same, flank timescales summed first & \multicolumn{2}{c}{60} \\
\midrule
\multicolumn{3}{l}{\emph{Slow term under corrections}} \\
$\tau_{\mathrm{fast}}$, daily cycle removed & 1.6 & 1.9 \\
$\tau_{\mathrm{slow}}$, daily cycle removed & 67 & 68 \\
ratio, daily cycle removed & \multicolumn{2}{c}{about 40} \\
$\tau_{\mathrm{slow}}$, 6-hour baseline & 38 & -- \\
\bottomrule
\end{tabular}
\end{table}

\subsection{Controlling for Time of Day}\label{sec:cycle}

Gas demand on Robinhood Chain has a strong daily pattern, and bursts concentrate in United
States daytime hours (Section~\ref{sec:when}). So, part of the shoulder may be due to the time
of day. To isolate the burst's own build-up from the time of day, we compare the median
alignment window with a \emph{control profile}: the median of the same twelve hours taken
exactly 24 hours before each peak (second row of Table~\ref{tab:profile}). All 654 bursts have
a control window. A control window can itself contain other bursts that occur at the same hour
of the day.

Two results follow from analyzing the control profile. First, at most 38\% of the slow rise can be attributed to the daily cycle. In the control profile, the usage rises by 0.23 from six hours to five minutes before the peak. In contrast, the rise is 0.61 in the burst profile over the same offsets. This is an upper bound because the control profile also carries the bursts that repeat at the same hour. Second, correcting for the control profile and refitting the data changes the slow timescales to 67 and 68 minutes (rise and fall, respectively). The ratio of the timescales is now about 40 (Table~\ref{tab:fit}). A slow term remains even after applying the controls. Its timescale is about a third as long as before.

\subsection{Phases in the Burst}

We now split the median alignment window into four phases by marking the five-minute mark on either side of the peak as the transition points. The four phases are named build-up, rise, fall, and decay (Table~\ref{tab:phases}). Each of these is named by the side of the peak and the term that governs it (Equation~\ref{eq:fuji}). At the five-minute mark, the slow term dominates (Appendix~\ref{app:det-shape}).
Two results follow from the phases. First, the fast exponential accounts for most of the change in the demand level during the rise and the fall phases. Similarly, the slow exponential accounts for most of the build-up and decay. Second, in individual bursts, the ascent and the descent take the same median time of 0.8 minutes. Moreover, the duration of the ascent phase predicts the duration of the descent phase better than peak usage or the pre-burst level (Table~\ref{tab:end}, Appendix~\ref{sec:end}).

\begin{table}[t]
\caption{The four phases of the median alignment window, split five minutes either side of the
peak. The share column is the fraction of the fitted change in level during the phase that
comes from the term in the governing-term column.}
\label{tab:phases}
\centering
\small
\begin{tabular}{llcr}
\toprule
phase & offsets from the peak & governing term & share \\
\midrule
build-up & before $-5$ min & slow & 86\% \\
rise & $-5$ to 0 min & fast & 97\% \\
fall & 0 to $+5$ min & fast & 96\% \\
decay & after $+5$ min & slow & 76\% \\
\bottomrule
\end{tabular}
\end{table}

To explain the origin of the shoulder, we compare the distribution of trading activity in the burst window with that in a random time period of equal length. In the 133 windows with a core, 96\% of the core's trading is in pools that also trade in the margin. A random block of minutes of the same length, drawn in the same window, gives 92.9\%. Heavily traded pools trade throughout the window. So, this comparison cannot tell where the shoulder comes from (Appendix~\ref{sec:one}). We defer the detailed analysis of the origins of the shape to future work.

\section{Composition of a Burst}\label{sec:comp}

In this section, we decompose the burst into what it is made of, which transactions consume the gas, and what the value of the trades inside is. The \emph{composition} of the burst summarizes the trading across all the non-routing pools (Section~\ref{sec:data}). Four quantities measure the composition: how many pools trade, the share of the largest pool, the share of the top ten, and the share of trading that is new in the last third of the window (compared to the initial third). The last quantity measures the \emph{rotation} in trading. If the burst is dominated by a single asset, then the burst has few pools, a top-heavy share, and little rotation. On the other hand, if the burst is due to a market-wide movement, then it has many pools, smaller shares, and high rotation.

Robinhood Chain has non-trading activity too. But the largest usage of gas in a burst is due to trading. Within the bursts, Uniswap swaps account for 37.6\% of the gas. The transfer and minting of tokens account for 30\% of the gas. The same events account for 42.1\% and 35.3\% outside the bursts (Table~\ref{tab:gas}, Appendix~\ref{app:det-comp}). A distinctive aspect of trading-heavy blockchains such as Robinhood Chain is that a significant fraction of transactions revert. On other chains, these are often submitted by bots that compete for arbitrage opportunities. Such transactions revert when the opportunity is captured by other bots, and still pay for the gas they used. The competition among these bots is known to generate congestion~\cite{daian20,wuoz26}. Within a burst, 17.1\% of the gas is spent on reverting transactions, compared to 8.5\% outside the burst time period.

\subsection{What Is in a Burst}
\begin{figure}[t]
\centering
\includegraphics[width=\columnwidth]{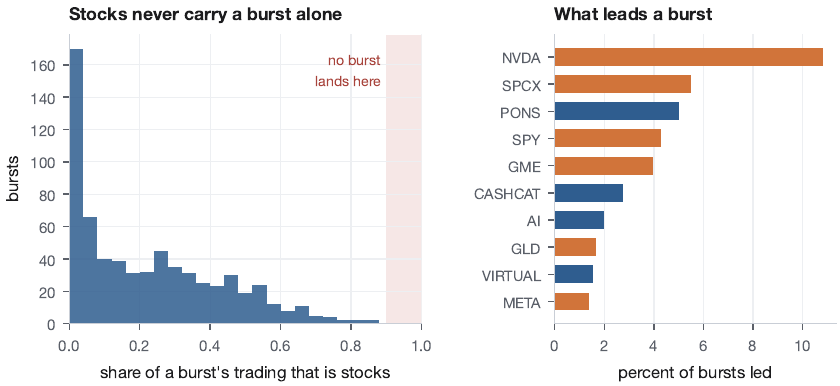}
\caption{Left: histogram of the share of trading in tokenized equities, one value per burst
interval (656 in all). The share does not exceed 90\% in any interval. Right: the ten tokens
that most often lead a burst, with the percentage of bursts each leads. Equities are in orange
and memecoins in blue.}
\label{fig:stocks}
\Description{Histogram of equity trading share per burst interval, and a bar chart of the ten most frequent leading tokens.}
\end{figure}

Table~\ref{tab:comp} summarizes the composition of a burst. The median burst window spans \num{2008} pools. Moreover, the composition narrows from \num{1282} pools in the median core to \num{674} pools in the median burst interval. The largest pool accounts for 7.6\% of the overall trading. This indicates that the trading is spread out across pools. The burst windows also exhibit high rotation: 43.7\% for the window and 33.3\% for the core. Tokenized equities and funds are the leading (by frequency of swaps) tokens in 34\% of the burst intervals. The NVDA token alone leads 10.8\% of the burst intervals. Figure~\ref{fig:stocks} shows a histogram of the share of stock tokens. There are no bursts where more than 90\% of the trading is related to stocks. Bursts led by equities are more frequent right after the New York open. During weekdays, more than half (54.8\%) of the bursts that start within 30 minutes of the stock market's opening time are led by equities (the overall percentage is 34\%). On the weekend, when the market is closed, the same comparison at the same clock time gives 34.5\% versus 33.8\%.

\begin{table}[t]
\caption{Composition of a burst: medians over the 200 burst windows, the 133 cores, and the
656 burst intervals. The first four rows are computed over the non-routing pools. The last row
is the share of swaps that pass through routing pools (Section~\ref{sec:trading}).}
\label{tab:comp}
\centering
\small
\begin{tabular}{lrrr}
\toprule
& window & core & interval \\
\midrule
distinct pools & \num{2008} & \num{1282} & 674 \\
largest pool's share & 0.076 & 0.113 & 0.144 \\
ten largest pools' share & 0.343 & 0.415 & 0.500 \\
late trading in pools absent at the start & 0.437 & 0.333 & 0.175 \\
routing pools' share of swaps & 0.163 & 0.200 & 0.275 \\
\bottomrule
\end{tabular}
\end{table}

We next ask how long a pool is active compared to its window. For each pool-window pair, we
measure the time between the first and third quartiles of the pool's swap times, as a fraction
of the window length. This value is 0.5 if the pool's swaps are spread evenly over the window.
Across the \num{68370} pairs with at least 50 swaps, the median is 0.23. Over all swaps in the
window, the fraction is 0.47, close to the 0.5 expected for an even spread. Therefore, a pool's
trading is about half ($0.23/0.47 \approx 0.5$) as spread out in time as the overall trading in
the window. This measure does not distinguish a pool that trades throughout from one that trades
in separate stretches (Appendix~\ref{app:det-comp}).

\subsection{Traders and Trading Frequency}

A burst trades about the same number of pools, but more than half of them differ from those in its control (Section~\ref{sec:data}). The number of pools rises by 1.05 times, while only 40\% of a burst's pools also traded in its control (Appendix~\ref{app:det-comp}). During a burst, the pools also trade harder: the number of swaps rises by a factor of 1.40. For each of the 150 sampled burst intervals, we calculate the ratio of swaps in the burst and its control. The median value of these ratios is 1.40. It includes the swaps in routing pools. The extra gas used by the bursts (1.68 times the control) can be attributed primarily to more senders and to each sender doing more transactions. As expected, the gas per transaction changes little because the gas used by a transaction does not depend on the demand. The excess is approximately the product of three ratios,
\begin{equation}
\underbrace{1.68}_{\text{gas}} \;\approx\;
\underbrace{1.21}_{\text{senders}} \times
\underbrace{1.26}_{\text{tx per sender}} \times
\underbrace{1.10}_{\text{gas per tx}} ,
\label{eq:decomp}
\end{equation}
where each factor is the median, over the same 150 pairs, of the ratio of that quantity during the burst to that in the corresponding control. For this decomposition, both the burst and its control are truncated to their central 600 blocks ($\sim$1 minute).
Here, senders are the addresses sending any transaction. This is a broader set than the traders defined in Section~\ref{sec:data} (Appendix~\ref{app:det-comp}). The counts come from the transaction receipts, not the swap logs, and they include transactions of any kind (swaps, transfers, reverts, and everything else). So the implied rise in transactions ($1.21 \times 1.26 = 1.52$) is not comparable to the rise in swaps above.

Only 15\% of the senders recur from a control to the following burst, compared to 17\% between two consecutive non-burst time periods. Overall, the chain's address population turns over by 80\% irrespective of the bursts. This shows that the traders on Robinhood Chain have high churn.

\subsection{Trading Value in a Burst}

The results above are based on the number of trades. They do not distinguish a burst with small-value trades from a burst with an equal number of higher-value trades. We define the value of a swap as the sum of the values of the inflow and outflow tokens, both converted to dollars based on the price of ether. We price only the version 3 swaps in pools with a quote asset, which are 93.5\% of the trades. Table~\ref{tab:value} shows the value of the swaps inside the burst.

\begin{table}[t]
\caption{Shares of tokens by value and by trades, as medians over the 656 burst intervals
(priced swaps only).}
\label{tab:value}
\centering
\begin{tabular}{lrr}
\toprule
& by value & by trades \\
\midrule
largest pool by value, its share & 0.176 & 0.063 \\
largest pool by trades, its share & 0.071 & 0.160 \\
ten largest pools, their share & 0.600 & 0.546 \\
late trading in pools absent at the start & 0.073 & 0.126 \\
\bottomrule
\end{tabular}
\end{table}

We draw attention to the following result from Table~\ref{tab:value}. The largest pool by value takes 17.6\% of the value against 6.3\% of the trades. The pool with the largest number of trades (16\%) accounts for 7.1\% of the overall value. Therefore, the share of the leading pool in a burst depends on whether the leading pool is selected by trades or by value.

\section{Models for the Arrival of Demand Bursts}\label{sec:when}

\begin{figure}[t]
\centering
\includegraphics[width=\columnwidth]{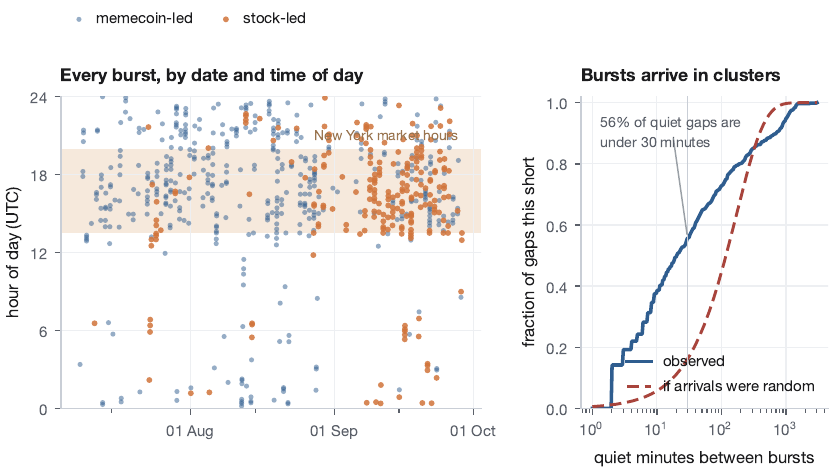}
\caption{Left: all 656 bursts by date and UTC time of day. Colors show whether an equity (orange)
or a memecoin (blue) led the trading. Right: the distribution of gaps between consecutive
bursts, compared with the distribution expected if bursts arrived independently at a constant
rate. Short gaps are more frequent than this distribution predicts.}
\label{fig:when}
\Description{Burst onset dates and UTC times colored by leading token class, beside the quiet-gap distribution and the independent-arrival benchmark.}
\end{figure}

In Section~\ref{sec:cycle}, at most 38\% of the slow rise can be attributed to the daily cycle in the trading. Neighboring bursts inside the alignment window may account for part of the rest. In the 24 bursts with no other burst within four hours of their onset, the slow timescale is smaller than that of other bursts (102 minutes versus 222 minutes, on the falling flank). Moreover, the peak is also lower at 2.26 times, compared to 2.56 times the usual. There are too few such bursts to test the difference (Appendix~\ref{sec:one}). In this section, we examine whether bursts arrive close together, and why.
Figure~\ref{fig:when} shows the arrival time of every burst. The bursts concentrate during United States daytime hours and cluster more than the daily cycle explains. Measuring the gaps between consecutive bursts, we find that more than half of them are shorter than 30 minutes. This is higher than the 15\% expected under independent arrivals and the 24\% expected under the daily-cycle assumption (Figure~\ref{fig:when}, right). The shorter gaps suggest alternative models for the arrival of bursts. We consider two such models in this section.

\subsection{Cascades or Self-Excitation}

Consider the sequence of onset times $t_1 < t_2 < \dots$ of the 656 bursts, where the onset is the first minute of the burst interval. Under the \emph{self-excitation} model, each burst raises the rate at which the next burst arrives. So the clusters of bursts are chains in which each burst induces the next. We write this as a Hawkes process~\cite{hawkes71,laub2024hawkes} with a periodic background,
\begin{equation}
\lambda(t) \;=\; \mu\,\phi\!\left(h(t)\right) \;+\; \sum_{t_i < t} \alpha\,e^{-\beta (t - t_i)} ,
\label{eq:hawkes}
\end{equation}
where $\lambda(t)$ is the rate of arrivals at time $t$, $\mu$ is the background rate measured in
bursts per hour, $h(t)$ is the hour of day, $\phi(\cdot)$ is a weighting factor assigned to each hour that averages to one over the day, $\alpha$ is the jump in rate caused by one burst, and
$1/\beta$ is the time over which that jump decays. Under this process, the ratio $\alpha/\beta$ is the expected number of bursts induced by a burst.

Under a \emph{cascade} model, a common external event leads to multiple independent bursts. These bursts do not induce each other directly. We follow Malmgren et al.~\cite{malmgren08}. The primary bursts
arrive as a Poisson process of rate $\mu_p\,\phi(h(t))$, where $\mu_p$ is the mean rate of
cascades per hour, and each primary burst is then followed by
\begin{equation}
N \sim \mathrm{Geometric}\!\left(\tfrac{1}{1+\nu}\right) - 1
\qquad\text{secondary onsets at rate } \rho ,
\label{eq:cascade}
\end{equation}
where the geometric distribution is on $\{1, 2, \dots\}$, $N$ is the number of secondary
bursts in the cascade, $\nu$ is its mean, and $\rho$ is the rate per hour at which the secondary bursts arrive.
In both models, we include the weighting factor $\phi$ to isolate the daily cycle from the rest (Appendix~\ref{app:det-when}).

We estimate the parameters of both models by minimizing the area between the distribution function of the gaps between bursts in the observed data and that of the gaps in simulations from the model. The objective function is
\begin{equation}
\begin{aligned}
A_r &= \int \left| F_{\mathrm{data}}(u) - F_r(u) \right| \, \mathrm{d}u ,\\
\bar{A} &= \frac{1}{R}\sum_{r=1}^{R} A_r , \qquad u = \ln \Delta t ,
\end{aligned}
\label{eq:area}
\end{equation}
where $\Delta t$ is a gap between consecutive arrivals, $F_{\mathrm{data}}$ is the empirical
distribution function of $u$ in the data, and $F_r$ is the same function for simulation
$r$ of $R$. For both models, the parameters are chosen to minimize the objective.

We draw the following three results from the fit. First, both models described here fit better than a simple daily-cycle model. The cascade model scores 0.268 and the self-excitation model scores 0.312, compared to 0.408 for the baseline (lower is better).
Second, the cascade model fits best. The fitted model opens about 2.7 cascades per day, each carrying on average 3.1 bursts including the one that opens it ($\nu \approx 2.1$). It produces 8.3 bursts per day, against the 8.0 observed between the first and the last burst.
Third, the self-excitation model cannot be excluded. The score difference between the two models is too small to discard the self-excitation model.

Finally, we test whether movements in the on-chain price of ether precede the arrival of bursts. In the ten minutes before a burst, the price moves 1.33 times as much as at the same hour on other days. The effect comes from bursts that follow another within an hour (1.45). For bursts that are more than two hours apart, the effect is not detectable (1.09, $p = 0.86$). Since a recent burst can itself move the price, we cannot identify the causal direction.

\section{Spread of a Burst}\label{sec:spread}
We now model the entry of traders into pools during a burst. The \emph{entry} is the earliest time a trader swaps in a pool in the twenty minutes around a burst window's peak (Section~\ref{sec:data}). We consider two models for the entry. Under a \emph{common trigger} model, the traders respond to a common external event. So the entry of traders during a burst is the cumulative result of individual responses to the event. This is our null model. Under an \emph{imitation} model, existing traders make later traders more likely to act. This is because trades are visible on a public ledger (Figure~\ref{fig:flow}). In this section, we show that the entry times of traders cluster beyond what the number of active traders predicts.

\subsection{Entry Models}

Let $S(t)$ be the total number of active traders in the pool at time $t$. Under the common trigger model, the number of traders who make their first trade at $t$, $E(t)$, is given by
\begin{equation}
E(t) \;\sim\; \mathrm{Binomial}\bigl(S(t),\, \theta(t)\bigr),
\label{eq:thin}
\end{equation}
where $\theta(t)$ is the fraction of the traders that are new. We estimate this as the sample average over 300 seconds, giving $\hat{\theta}(t)$. This is a memoryless model. Now consider the imitation model. Since each existing trader draws in more new traders, the time series of the residual $r(t)$ between the observed and the expected new entrants has autocorrelation. The residual is
\begin{equation}
r(t) \;=\; E(t) - \hat{\theta}(t)\, S(t)
\label{eq:resid}
\end{equation}

We measure the autocorrelation at lags ranging from one to forty seconds. As $\hat{\theta}(t)$ is estimated, the null model also has non-zero autocorrelation. We define \emph{excess} as the observed autocorrelation minus the mean autocorrelation of bootstrap samples from the null model (Appendix~\ref{app:det-spread}).

\subsection{Entries Cluster Beyond the Common Trigger}

\begin{figure}[t]
\centering
\includegraphics[width=\columnwidth]{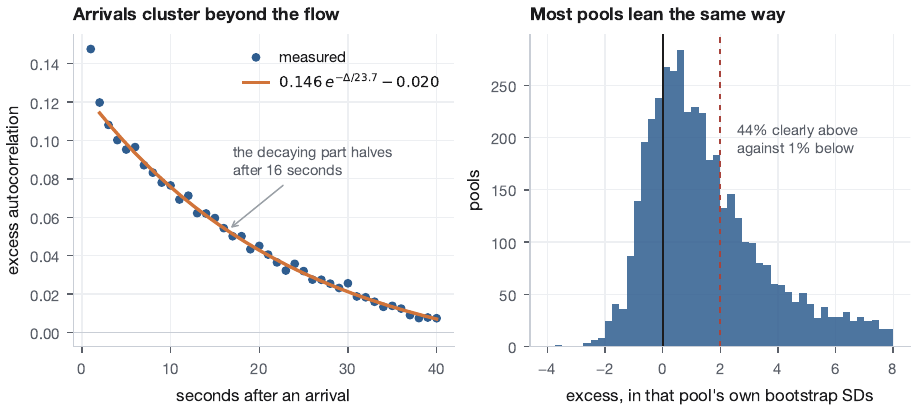}
\caption{Left: excess autocorrelation of entries by lag, with the fit of
Equation~\ref{eq:kernel}. The one-second lag is excluded from the fit. Right: distribution of
the excess over the \num{5144} pool-windows. Each value is the excess averaged over lags and
divided by the standard deviation of that pool-window's bootstrap samples. The dashed line
marks two.}
\label{fig:spread}
\Description{Excess entry autocorrelation decays with lag. The distribution across pool-windows is predominantly positive relative to each bootstrap benchmark.}
\end{figure}

Figure~\ref{fig:spread} shows the excess by lag (in seconds) and by pool-window. Over \num{5144} pool-windows extracted from the thirty largest pools in the 200 burst windows, the average excess is $+0.050$. The excess is largest at the shortest lag (1 second) and decays smoothly. This is also observed individually for most pool-windows (Appendix~\ref{app:det-spread}).

The decay in excess can be described by the following exponential function,
\begin{equation}
\mathrm{excess}(\Delta) \;=\; A_{\mathrm{k}}\,e^{-\Delta/\tau_{\mathrm{k}}} + c_{\mathrm{k}} ,
\label{eq:kernel}
\end{equation}
where $\mathrm{excess}(\Delta)$ is the mean excess at a lag of $\Delta$ seconds,
$A_{\mathrm{k}}$ is the amplitude of the decaying component,
$\tau_{\mathrm{k}}$ is the decay time, and $c_{\mathrm{k}}$ is the value to which the excess converges.
Fitting the data by nonlinear least squares gives $A_{\mathrm{k}} = 0.146$, $\tau_{\mathrm{k}} = 23.7$ seconds, and
$c_{\mathrm{k}} = -0.020$. The negative value of $c_{\mathrm{k}}$ is the result of fitting a single exponential at finite lag. The interquartile range of $\tau_{\mathrm{k}}$ is [23.1, 24.3] across samples of pool-windows.

We read the following two results from the fit. First, the decay over 24 seconds does not match either of the two other timescales: the block time (milliseconds) and the demand profile (minutes) (Section~\ref{sec:shape}). However, it is of the same order as the time a person takes to act on new market information. In futures markets, the median time taken by a manual trader between an execution and their next order exceeds nine seconds. The delay exceeds fifteen seconds when traders are weighted equally~\cite{fishe2016}. Second, the excess grows with the swap count but not with the number of traders. We observe this by sorting the pool-windows. When sorted by swap count, the excess rises from 0.006 to 0.115 between the smallest and the largest quarters. Sorting by the number of distinct traders does not exhibit such monotonic behavior (0.039, 0.120, 0.026, and 0.014 across the quarters).

\subsection{Trader Timing}

We now study when traders enter pools. A trader's \emph{entry position} is how early the trader first appears in a pool, measured as a fraction of the pool's active span in that window, from its first swap to its last. We define \emph{earliness} as the pool-window's mean position minus the trader's entry position, adjusted for the trader's swap count (Appendix~\ref{app:timing}). A positive earliness means that the trader enters earlier than average.
We make the following two observations. First, the earliness of traders repeats. Splitting the pool-windows into two halves, we observe that a trader's earliness correlates at $+0.311$ across the two halves. The correlation is $+0.001$ if the trader labels are shuffled. We do not observe a large correlation between earliness and the number of pools a trader touches ($+0.063$) or the total number of swaps ($-0.004$). Second, we observe a stable directed ordering among the active traders. Among the 600 most active traders, we compute the difference in earliness for each pair of traders after removing each trader's mean earliness. This measure correlates at $+0.377$ across two random halves versus $+0.007$ if the labels are shuffled. We do not read this result as establishing that one trader observes the other. A stable ordering is also possible between traders with different context-dependent habits.

\section{Related Work}\label{sec:rel}

\textbf{Demand and congestion pricing.}
Existing work has analyzed congestion pricing on different platforms. On blockchains, Roughgarden~\cite{roughgarden21} studies attainable guarantees for transaction fee mechanisms. Chung and Shi~\cite{chungshi23} extend the analysis to coalitions of users and miners. Reijsbergen et al.~\cite{reijsbergen21} measure how block size oscillates, Donmez and Karaivanov~\cite{donmez21} relate fees to the utilization of chain resources, and
Liu et al.~\cite{liu22} measure fee dispersion, waiting times, and the impact on consensus. On other platforms, Kelly~\cite{kelly97} relates network charges to allocated rates. Castillo et al.~\cite{castillo17} study congestion and surge pricing on ride-hailing services. Chen and Sheldon~\cite{chen16} measure how drivers respond to surge pricing.
In all of these works, the demand burst is an input to the pricing. In this work, we model the shape, breadth, and arrival of the demand.

\textbf{Bursts and their arrival.}
Kleinberg's seminal work~\cite{kleinberg02} proposes a formal model of bursts of topics in email and document streams.
Crane and Sornette~\cite{crane08} separate bursts with a sudden peak from bursts with a slow rise before the peak. They attribute the slow rise to endogenous build-up. Yang and Leskovec~\cite{yang11} cluster online activity by its shape around the peak.
The median profile in our data resembles a slow rise followed by a rapid ascent, although we do not establish that the rise is endogenous.
Barab{\'a}si~\cite{barabasi05} models heavy-tailed intervals with priority queues, and V{\'a}zquez et al.~\cite{vazquez06} extend that model. Malmgren et al.~\cite{malmgren08} reproduce heavy-tailed distributions using periodic Poisson cascades. We use this construction and distributional distances to infer the model parameters. In Karsai et al.~\cite{karsai12}, the authors separate burstiness from temporal correlation. Jo et al.~\cite{jo12} show how daily cycles can inflate the measured burstiness. We test this distinction in Section~\ref{sec:when}.

\textbf{Diffusion and attention.}
Bikhchandani et al.~\cite{bikhchandani92} propose a model for information cascades, where imitation outweighs private signals. The cascades of Section~\ref{sec:when} are a different construction and do not include imitation. We find clustering consistent with a similar mechanism in Section~\ref{sec:spread}. Leskovec et al.~\cite{memetracking09} follow phrases through news episodes. We take a similar approach to follow pools through burst time periods.
Weng et al.~\cite{weng13} model the spread of memes through online communities.
Wang and McCallum~\cite{tot06} model continuous changes in topic importance.
On-chain data lacks the network and content that these approaches use. Therefore, we only test time dependence between the traders.
Wu and Huberman~\cite{wu07} measure fading attention to news.
Salganik et al.~\cite{salganik06} show that social information affects music choices and rankings.
The token rotation measure we use in the paper measures trading without explicitly observing attention or exposure to social feeds.

\textbf{Automated market makers.} Angeris et al.~\cite{angeris20} analyze when on-chain market prices track an external reference price. Sch{\"a}r~\cite{schar21} is a survey of blockchain-based markets.
Daian et al.~\cite{daian20} show how the competition among bots generates congestion on blockchain resources.
More recently, Wu and {\"O}z~\cite{wuoz26} find that bots that submit speculative transactions to capture arbitrage opportunities use a significant share of the gas.
The elevated share of gas spent on reverting transactions during bursts (Section~\ref{sec:comp}) is consistent with the analysis.

\section{Discussion}\label{sec:disc}

Robinhood Chain has quickly become a venue for the trading of tokenized equities, funds, and memecoins.
This trading is predominantly on the Uniswap application. Beyond the on-chain activity, the trades are also visible on social feeds that show what others are buying. The chain consistently experiences bursts of activity (about eight times a day). The demand burst on this venue has the shape of a neck of minutes on a shoulder of hours. Part of the shoulder is the daily cycle: at most 38\% of the slow rise before the peak is attributable to it. Within a burst, the number of pools is about the same as in a quiet period, but more than half of them differ. There are more swaps, and the gas spent on reverting transactions doubles.
The leading pool by value holds about a sixth of a burst's value, and the leading pool by trades about a sixth of its trades.
Bursts cluster beyond the daily-cycle pattern, and a cascade model fits the arrival gaps better than a self-excitation model, although the self-excitation model cannot be excluded.
Within a burst, the arrivals of traders cluster beyond what a common trigger model suggests. The excess over the common trigger model decays over 24 seconds and is larger in pools with more swaps.

From the chain's perspective, the shape analysis suggests what its (gas) capacity has to absorb and for how long. The peak of the median burst is 2.5 times the usual level, and the neck lasts for five minutes on each side of the peak. The shoulder period of a burst is about 1.3 times the usual level and lasts for two hours on either side of the peak. Overall, 12.2\% of the gas is consumed in the bursty time periods, which make up 6.1\% of the time. The gas prices on blockchains are set dynamically and are primarily based on the recent usage. The shape suggests that a pricing mechanism may need components that respond to both the neck and the shoulder of the burst.
The modeling of burst arrivals indicates what can be anticipated on the chain. The arrivals of bursts are consistent with factors beyond the burst, and self-excitation cannot be ruled out. In addition, the arrivals of traders are also clustered. These are the quantities a pricing rule for gas would need to take into account.

\subsection{Limitations}

The data in this analysis was collected from one chain over its first three months of operation. We focus on the Uniswap application and the token swaps that happen on this platform. The burst-to-control ratios in this work are based on a sample of burst intervals. In the shape analysis, the slow timescale depends most on the detection baseline. The slow timescale falls from 218 minutes to 38 minutes when using a six-hour baseline. The sampling uncertainty of the margin between the cascade and self-excitation models is not estimated. The null model counts only the first swap of each trader. Repeated swaps by the same trader keep it among the active traders, which can produce clustering without imitation. The imitation model among the traders is not causally identified~\cite{aral09,shalizi11}.

\bibliographystyle{splncs04}
\bibliography{refs}

\appendix
\section{Measurement Details}\label{app:details}

\subsection{Data and Definitions}\label{app:det-data}

Traders are identified from the transaction's \texttt{from} field. Across the 13.7 million
swaps in the twenty minutes around the peak of each of the 200 burst windows, there are
\num{732619} different traders. The largest trader sends 0.7\% of swaps.

The thresholds used in Equations~\ref{eq:def} and~\ref{eq:window} were chosen by inspection.
We keep an alignment window when at least 648 of its 721 minutes lie inside the data. Two bursts at the end were removed because of this restriction.

\subsection{Shape and Descent}\label{app:det-shape}

\begin{figure}[t]
\centering
\includegraphics[width=\columnwidth]{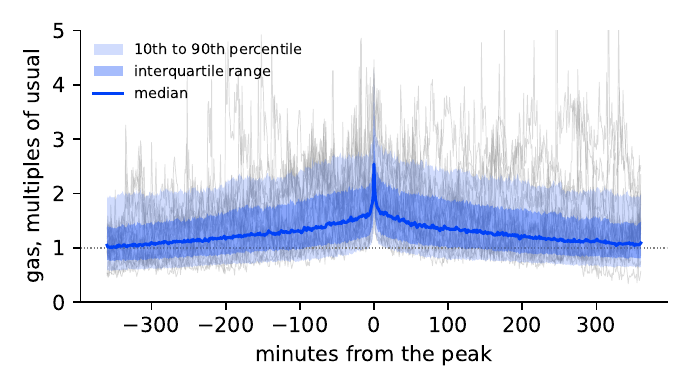}
\caption{The 654 alignment windows, in multiples of the usual level: median, interquartile
range, 10th to 90th percentile range, and fifteen individual windows drawn faintly. The whole
distribution rises toward the peak.}
\label{fig:aligned}
\Description{Percentile bands of gas around the median burst profile over twelve hours, with a few individual windows.}
\end{figure}

Figure~\ref{fig:aligned} shows the spread of the 654 aligned windows around the median profile (Section~\ref{sec:shape}).

For a robust measure of the fast and slow timescales, we resampled the 654 bursts and refit the median profile of each resample.
The interquartile range for $\tau_{\mathrm{fast}}$ is 2.2 to 3.6 minutes, and for $\tau_{\mathrm{slow}}$ it is 198 to 282 minutes. On the falling flank, these ranges are 3.3 to 9.8 minutes and 185 to 294 minutes for the fast and slow exponentials, respectively. The slow term has heavy upper tails, and it is unidentified in nearly 6\% of the resamples on the rise. The fast term is relatively less certain on the descent.
The ratios reported in Table~\ref{tab:fit} are computed from the unrounded fit, so they differ slightly from the ratios of the rounded values in the table. Under the 6-hour baseline, the slow timescale is 38 minutes (versus 218 minutes in the 24-hour baseline). The fast timescale is insensitive to the baseline, so we do not report it.
We exclude the peak minute from the two-exponential fit. The model reaches 2.16 at the peak, against 2.54 in the data.
At the five-minute offset, the fast term retains 16\% of its value at the peak on the rise and 31\% on the fall.

\textbf{Descent.}\label{sec:end}

Gas usage is binned into ten-second intervals. A smooth gas series is obtained by averaging the usage over a minute.
The peak is the largest bin inside the burst. The pre-burst level is the median over the 30 minutes before the peak.
Half-height is the midpoint between the pre-burst level and the peak. The ascent is the time from the last time the level is below half-height to the peak. Similarly, the descent is the time from the peak to the first time the level falls below half-height. The 654 bursts in the data each have three hours of gas usage data on either side of the peak and have both ascent and descent phases.
Since the ascent and the descent are both measured from the same half-height, an error in the peak or the pre-burst level moves both durations in the same direction. Therefore, the correlations in Table~\ref{tab:end} partly reflect how the duration is measured.

\begin{table}[t]
\caption{Spearman correlations between the duration of a burst's descent and three predictors,
computed over 654 bursts. Peak height and pre-burst level both set the half-height threshold, so
their correlations with the descent partly reflect how it is measured.}
\label{tab:end}
\centering
\begin{tabular}{lrr}
\toprule
& Spearman $\rho_s$ & $p$ \\
\midrule
height of the peak & $+0.157$ & $5\times10^{-5}$ \\
pre-burst level & $-0.196$ & $5\times10^{-7}$ \\
duration of the ascent & $+0.428$ & $<10^{-10}$ \\
\bottomrule
\end{tabular}
\end{table}

\subsection{Composition}\label{app:det-comp}

The gas used by different transaction types is based on receipts from a sample of \num{6000} blocks each from inside and outside the bursts (Table~\ref{tab:gas}).
We manually classify the tokens as equities or memecoins based on their tickers.

\begin{table}[t]
\caption{Share of gas by transaction type, inside and outside the burst intervals. Each column
is computed over a sample of \num{6000} blocks. The share of gas spent on reverted
transactions doubles inside the bursts.}
\label{tab:gas}
\centering
\small
\begin{tabular}{lrr}
\toprule
transaction kind & inside bursts & outside \\
\midrule
emits a Uniswap swap & 37.6\% & 42.1\% \\
transfers or mints only & 30.0\% & 35.3\% \\
reverted & 17.1\% & 8.5\% \\
emits other events & 6.5\% & 8.9\% \\
emits no events & 8.9\% & 5.2\% \\
\bottomrule
\end{tabular}
\end{table}

Over the 150 burst-control pairs, the median Jaccard index (intersection over union) of the pool sets is 0.25. The pools in the intersection account for 35\% of the swaps in the burst.

The largest pool's median share of 14.4\% in the burst intervals is a lower bound on the leading token's share (Table~\ref{tab:comp}). On the priced sample, Table~\ref{tab:value} shows that the share of the leading pool by trades is 16.0\%.

Swap value is estimated as inflow plus outflow, using the on-chain ether price and valuing stablecoins at their target price.
Both columns of Table~\ref{tab:value} use the same pools and minutes. Row one selects the leader by value, and row two selects it by number of trades. Row three selects the ten largest by the measure of each column.

The largest pool by trades and the largest pool by value are the same in 36\% of the bursts. The ten largest pools by value account for 60\% of the value, and the ten largest by trades account for 54.6\% of the trades. The third column of Table~\ref{tab:comp} selects pools by trades
and includes all swaps.

The product in Equation~\ref{eq:decomp} holds only approximately because each factor is a median of separate ratios.

\subsection{Arrival}\label{app:det-when}

To model arrivals, we use models with multiple independent agents. Single-agent models such as that of V{\'a}zquez et al.~\cite{vazquez06} are not considered in this study.
The cascade model generates each cascade independently. So cascades can overlap in time. We use two-level cascades: a secondary burst does not generate further bursts.
Both arrival models have the same number of parameters. Therefore, we do not adjust the fit for model complexity.

We used Nelder--Mead search on $\bar{A}$ with $R = 6$ to find the parameters. The values reported in the paper are mean estimates over 25 simulations at the fitted parameters. The margin between the two models is 0.043, with a standard error of 0.009. This standard error reflects only the randomness of the simulation. We do not estimate the sampling uncertainty of the margin. Therefore, we do not exclude the self-excitation model. The weighting function $\phi$ is estimated from hourly arrival counts. Fitted by maximum likelihood, the branching ratio with the estimated $\phi$ is 0.48, compared to 0.66 with a uniform weighting function. The self-excitation model fitted in Section~\ref{sec:when} has a branching ratio of 0.69 and a decay time of 101 minutes.

The ether price is read directly from the pool state. The price movement in the ten minutes before a burst is 1.33 times the movement at the same hour on other days. It is 1.22 times in the window from 60 to 30 minutes before. For the ten-minute window, the factor is 1.45 ($p = 10^{-4}$) for the 410 bursts that occur within an hour of the previous burst. But it is 1.09 ($p = 0.86$) for the 169 bursts that occur at least two hours after the previous one. Bursts between one and two hours after the preceding burst enter neither group.

\subsection{Spread}\label{app:det-spread}

We only consider pool-window pairs that have at least 200 swaps, 50 distinct traders, and 300 one-second bins.
This leaves \num{5144} out of the \num{6000} candidate pairs. The mean autocorrelation of the bootstrap samples is $+0.007$ at one second and $+0.002$ at 40 seconds, with an average of $+0.004$. We subtract this to obtain the reported excess.
The excess is positive in 80\% of the pool-windows. In 44.3\% of the pairs, it is more than two standard deviations.

Variations faster than the 300-second smoothing can leave positive autocorrelation unaccounted for by the null model.
Conditioning on $S(t)$ can also absorb imitation that changes the flow of swaps.
Resampling the \num{5144} pool-windows \num{1000} times and refitting gives an interquartile
range of 23.1 to 24.3 seconds and a 5th to 95th percentile range of 22.3 to 25.2 seconds.

\section{The Shoulder Test}\label{sec:one}\label{app:det-one}

Each of the 133 burst windows with a core is split into the core, which is the minutes in burst intervals, and the margin, which is the remaining minutes (Equation~\ref{eq:core}). If the spike and the shoulder are the same event, then the margin and the core trade in the same pools. In the median window, 96\% of the core's trading is in pools that also trade in the margin, compared to 92.9\% for a random block of the same length. Moreover, the core's overlap with the margin exceeds that of a random block in 87\% of the windows, by a median of 0.02. Since a random block shares nearly as much trading with the margin, the test is not powerful enough to distinguish a shoulder of the current burst from that of a previous burst. Two further limits apply to the identification. First, the random block is contiguous, while a core may span disjoint time periods. Second, the margin covers only the part of the shoulder that is above the window threshold.

The isolated bursts of Section~\ref{sec:when} are the 24 bursts whose onset is more than four hours from the onsets of both the previous and the next burst.
In these bursts, the slow timescale is 102 minutes, compared to 222 minutes for the remaining bursts. The peak is 2.26 against 2.56 times the usual level. For the isolated bursts, Equation~\ref{eq:fuji} is fitted only on the falling flank and without the daily-cycle correction. The slow timescale changes from 199.2 to 222 minutes after removing the isolated bursts. With only 24 bursts and one flank, the isolated estimate of 102 minutes is uncertain.

\section{Trader Timing Checks}\label{app:timing}

The entry position is defined as the fraction of the pool's active span at which the trader first appears.

For trader $s$ in pool-window $w$, the adjustment is
\begin{equation}
q_{sw}-\bar q_w=a+b\log_{10}n_{sw}+\varepsilon_{sw},
\label{eq:speed}
\end{equation}
where $q_{sw}$ is the entry position, $\bar q_w$ is the average entry position over all traders that appear in $w$, and $n_{sw}$
is the trader's swap count in $w$.
The coefficients $a$ and $b$ are fitted by pooled ordinary least squares.

Earliness in Section~\ref{sec:spread} is $-\hat\varepsilon_{sw}$, so a positive value means an
entry earlier than the window's mean. The split-half comparison covers \num{123285} traders
observed in \num{2708314} (trader, pool, window) triples.
For the comparison, each trader's pool-windows are split at random into two halves.
The correlation compares each trader's mean earliness across these two halves.

The pair ordering is the average difference in
$\tilde\varepsilon_{sw}=\hat\varepsilon_{sw}-\overline{\hat\varepsilon}_s$
between two traders over their common pool-windows. The 600 most active traders yield \num{157160} pairs with at least ten common pool-windows.
The correlation between the halves is $+0.377$, compared with $+0.007$ after shuffling the labels.
Splitting the pairs so that the halves use disjoint windows gives a correlation of $+0.346$.
Splitting by disjoint pools gives a correlation of $+0.147$ over about \num{149000} pairs. Splitting by time, the correlation between the earlier and the later halves is $+0.278$.

\end{document}